\documentclass[prl,aps,twocolumn,showpacs,nobibnotes,epsf]{Revtex4}

\usepackage{graphicx}
\usepackage{dcolumn}
\usepackage{bm}
\usepackage{SIunits}
\usepackage{verbatim}
\usepackage{placeins}

\begin{document}
\title{Magnetism and superconductivity in single crystals $Eu_{1-x}Sr_xFe_{2-y}Co_{y}As_2$}
\author{Q. J. Zheng, Y. He, T. Wu, G. Wu, H. Chen, J. J. Ying, R. H.~Liu, X. F. Wang, Y. L. Xie, Y. J. Yan, Q. J. Li}
\author{X. H. Chen}
\altaffiliation{Corresponding author} \email{chenxh@ustc.edu.cn}
\affiliation{Hefei National Laboratory for Physical Science at
Microscale and Department of Physics, University of Science and
Technology of China, Hefei, Anhui 230026, People's Republic of
China\\}

\begin{abstract}
We systematically studied the transport properties of single
crystals of $Eu_{1-x}Sr_xFe_{2-y}$Co$_{y}As_2$. Co doping can
suppress the spin-density wave (SDW) ordering and induces a
superconducting transition, but a resistivity reentrance due to the
antiferromagnetic ordering of $Eu^{2+}$ spins is observed,
indicating the competition between antiferromagnetism (AFM) and
superconductivity. It is striking that the resistivity reentrance
can be completely suppressed by external magnetic field (H) because
a metamagnetic transition from antiferromagnetism to ferromagnetism
for $Eu^{2+}$ spins is induced by magnetic field. Superconductivity
without resistivity reentrance shows up  by partial substitution of
Eu$^{2+}$ with non-magnetic Sr$^{2+}$ to completely destroy the AFM
ordering of $Eu^{2+}$ spins. These results suggest that the
antiferromagnetism destroys the superconductivity, while the
ferromagnetism can coexist with the superconductivity in the
iron-based high-$T_c$ superconductors.
\end{abstract}


\vskip 300 pt

\maketitle

Interplay between magnetism and superconductivity has long been an
interesting issue in condensed matter physics, where in most cases,
these two orders compete with each other. Theoretical work done by
Ginzberg, Baltensperger and Straesler predicted that long range
ferromagnetism would greatly damage superconductivity while
antiferromagnetism could coexist with superconductivity to some
extend\cite{Ginzberg, Baltensperger}. Indeed, in traditional
magnetic superconductors like RMo$_6$O$_8$ and RRh$_4$B$_4$ (R =
magnetic rare earth ions)\cite{Fertig, Ishikawa}, superconductivity
is found to be moderately robust coexisting with antiferromagnetism,
yet fragile with ferromagnetism (FM). However, coexistence of
ferromagnetism and superconductivity was observed in heavy fermion
system UGe$_2$ and unconventional high-T$_c$ cuprates superconductor
RuSr$_2$GdCu$_2$O$_8$\cite{Bernhard}, where local moments locate far
from conducting plane. Apart from all mentioned above, one of the
most typical families of magnetic superconductor is RNi$_2$B$_2$C
where R = Ho, Er, Tm, Yb, Lu\cite{Eisaki, Canfield} with
ThCr$_2$Si$_2$-type structure. In these materials, a resistivity
reentrance below superconducting transition temperature was observed
due to the antiferromagnetic ordering of R ions. Therefore, the
interaction between superconductivity and magnetic ordering is still
puzzling. Here we report intriguing results that the
antiferromagnetism destroys the superconductivity, while the
ferromagnetism can coexist with the superconductivity in the
iron-based high-$T_c$ superconductors. The antiferromagnetism and
ferromagnetism can be manipulated by tiny magnetic field. These
results are definitely significant to understand the interaction
between superconductivity and magnetic ordering.

\begin{figure}[htbp]
\centering
\includegraphics[width=0.45\textwidth]{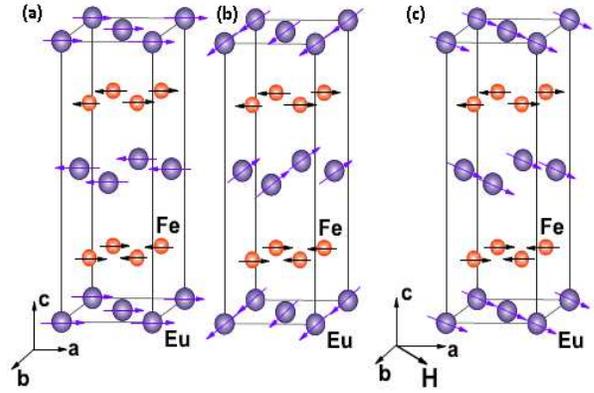}
\caption{(Color online). (a) and (b): Two possible collinear
magnetic structures of EuFe$_2$As$_2$ at zero field; (c): External
magnetic field (H) induced ferromagnetic structure of
EuFe$_2$As$_2$. Figure reproduced from [16].}\label{fig1}
\end{figure}

Since Fe ions are more likely to convey magnetic moment in various
ways due to their nearly half filled 3d orbital and larger degree of
freedom in electronic spin states than cuperates, the new iron-based
family of high Tc superconductors\cite{Kamihara, Chen} is well
expected to be a promising category among which might exist new
magnetic superconductors. In fact, an AFM SDW ordering of $Fe^{2+}$
is widely observed in the parent compounds of iron-based
superconductors.\cite{Cruz,huang} The $ARFe_2As_2$ (AR=Ba, Sr, Ca,
Eu) 122 family have a relatively simply-structured charge reservoir
layer which may serve as an ideal candidate to embed magnetism into
this system. The $Ba^{2+}$, $Sr^{2+}$ and $Ca^{2+}$ are non-magnetic
ions, while $Eu^{2+}$ is magnetic ion with s=7/2. The magnetic
ordering of $Eu^{2+}$ spin occurs at about 17 K. As illustrated in
Fig 1, an A-type antiferromagnetic structure for Eu$^{2+}$
sublattice is proposed by Wu et al. and Jiang et al.\cite{Wu, Jiang}
in parent compound EuFe$_2$As$_2$, where local moments align
collinearly to form strong FM order within ab plane and weak AFM
order along c axis (Fig.1a and 1b). The interlayer AFM coupling can
be tuned to FM coupling (Fig.1c) by a tiny external magnetic field.
In Ba122 system, the superconductivity can be induced by
substitution of K for Ba or Co-doping on Fe
site.\cite{Rotter,Chend,Wangb} Therefore, $EuFe_2As_2$ is a good
system to study the interaction between superconductivity and
magnetic ordering of $Eu^{2+}$ spins. Therefore, we systematically
measure the transport properties to study the interaction between
superconductivity and magnetism in the Co-doped Eu122 system in
which the magnetic ordering for Eu$^{2+}$ sublattice nearly does not
change with Co doping. Then, we substituted Eu with Sr to
systematically suppress the magnetic ordering and study the effect
of suppression of magnetic ordering on superconductivity. It is
found that a resistivity reentrance due to the antiferromagnetic
ordering of $Eu^{2+}$ spins is observed below $T_c$, indicating the
competition between antiferromagnetism (AFM) and superconductivity.
It is striking that the resistivity reentrance can be completely
suppressed by external magnetic field (H) because of the
metamagnetic transition from antiferromagnetism to ferromagnetism
for $Eu^{2+}$ spins induced by H. These results suggest that the
antiferromagnetism destroys the superconductivity, while the
ferromagnetism favors the superconductivity in the iron-based
high-$T_c$ superconductors.

\begin{figure}[htbp]
\centering
\includegraphics[width=0.5\textwidth]{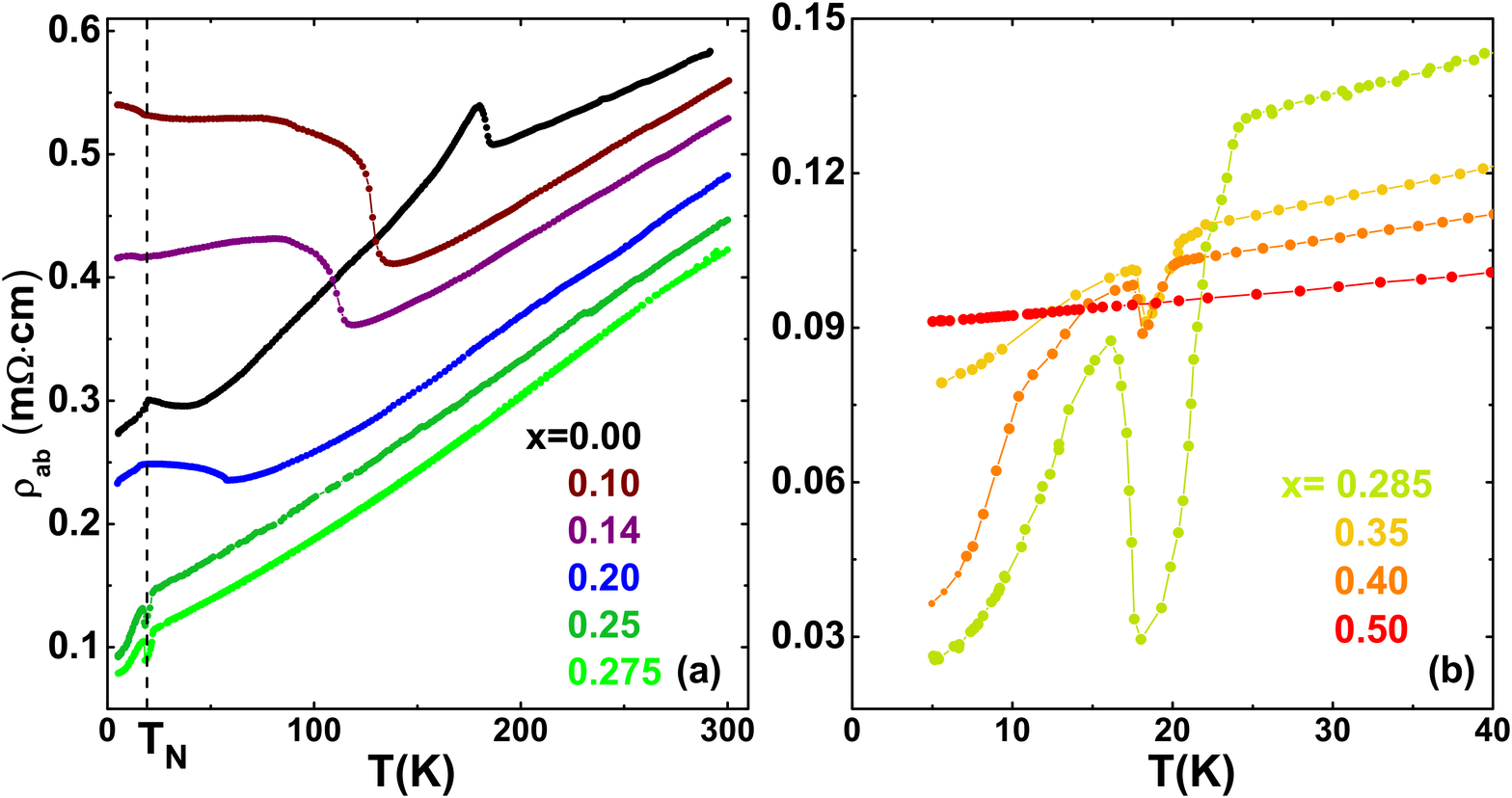}
\caption{(Color online). Temperature dependence of in-plane
resistivity for the EuFe$_{2-x}$Co$_{x}$As$_2$ single crystals. (a):
$x$=0, 0.1, 0.14, 0.2, 0.25 and 0.275 from 5 K to 300 K; (b):
$x$=0.275, 0.285, 0.35, 0.4 and 0.5 from 5 K to 40 K. Note that all
the dips in resistivity correspond to the same temperature as T$_N$
in parent EuFe$_2$As$_2$.}\label{fig2}
\end{figure}

\begin{figure}[htbp]
\centering
\includegraphics[width=0.5\textwidth]{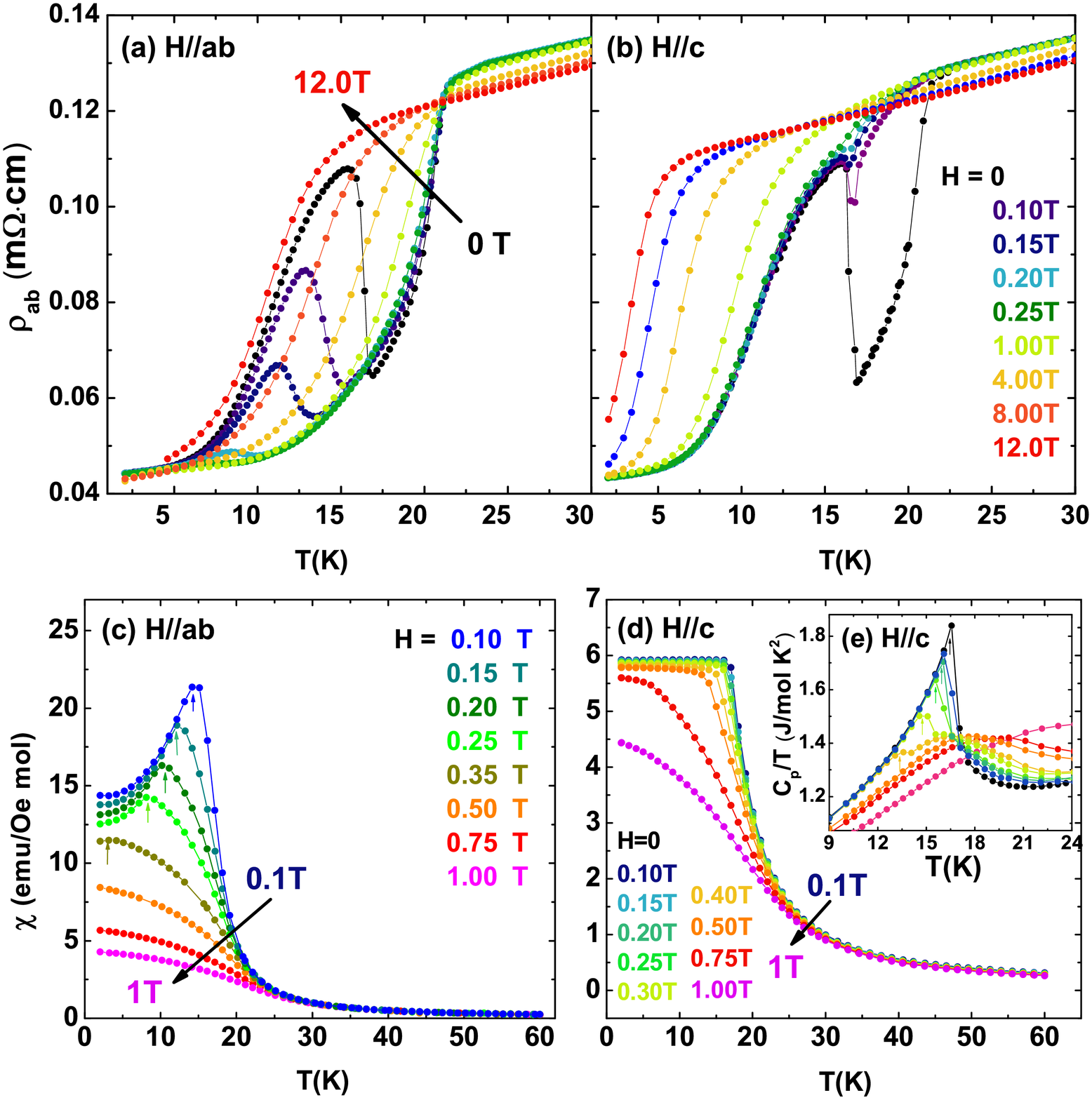}
\caption{(Color online). Temperature dependence of resistivity and
susceptibility under different magnetic fields for optimal doped
single crystal EuFe$_{1.715}$Co$_{0.285}$As$_2$. (a): H applied
within ab-plane; (b): H applied along c-axis; (c): H applied within
ab-plane; (d): H applied along c-axis; (e): Specific heat under
different H applied along c-axis (enlarged from 9 K to 24
K).}\label{fig3}
\end{figure}

The single crystals of EuFe$_{2-x}$Co$_{x}$As$_2$ and
Eu$_{y}$Sr$_{1-y}$Fe$_{2-x}$Co$_{x}$As$_2$ were synthesized via
conventional self-flux method\cite{Wang}. Figure 2 demonstrates the
systematic evolution in temperature dependent resistivity in the
temperature range from 290K down to 5K for
EuFe$_{2-x}$Co$_{x}$As$_2$ single crystals. The left figure (Fig.2a)
presents the resistivity data for the underdoped samples, and the
right figure (Fig.2b) illustrates resistivity as a function of
temperature for the optimal doped ($x$=0.285) and overdoped samples.
In underdoped region, the resistive behavior largely resembles that
of BaFe$_{2-x}$Co$_{x}$As$_2$ in terms of high temperature
features\cite{Wangb}. The only difference happens at about 17 K,
where a dopant independent kink emerges, corresponding to the AFM
ordering established among Eu$^{2+}$ layers. Further Co-doping
introduces a superconducting transition in resistivity around 22 K,
but consequently a resistance reentrance shows up just below the AFM
ordering temperature, so that no zero resistivity is observed with
the temperature down to 4.2 K in EuFe$_{2-x}$Co$_{x}$As$_2$ system.
This is totally different from that observed in
BaFe$_{2-x}$Co$_{x}$As$_2$ system. It suggests that the introduction
of Co$^{2+}$ into FeAs layer fails to bring about superconductivity
in EuFe$_{2-x}$Co$_{x}$As$_2$ system. Even the optimally doped
single crystal at x = 0.285 just shows a resistivity decrease by
less than 80\% before the occurrence of resistivity reentrance. No
superconducting transition and no resistivity reentrance are
observed in the overdoped crystal with x=0.5.

\begin{figure}[htbp]
\centering
\includegraphics[width=0.45\textwidth]{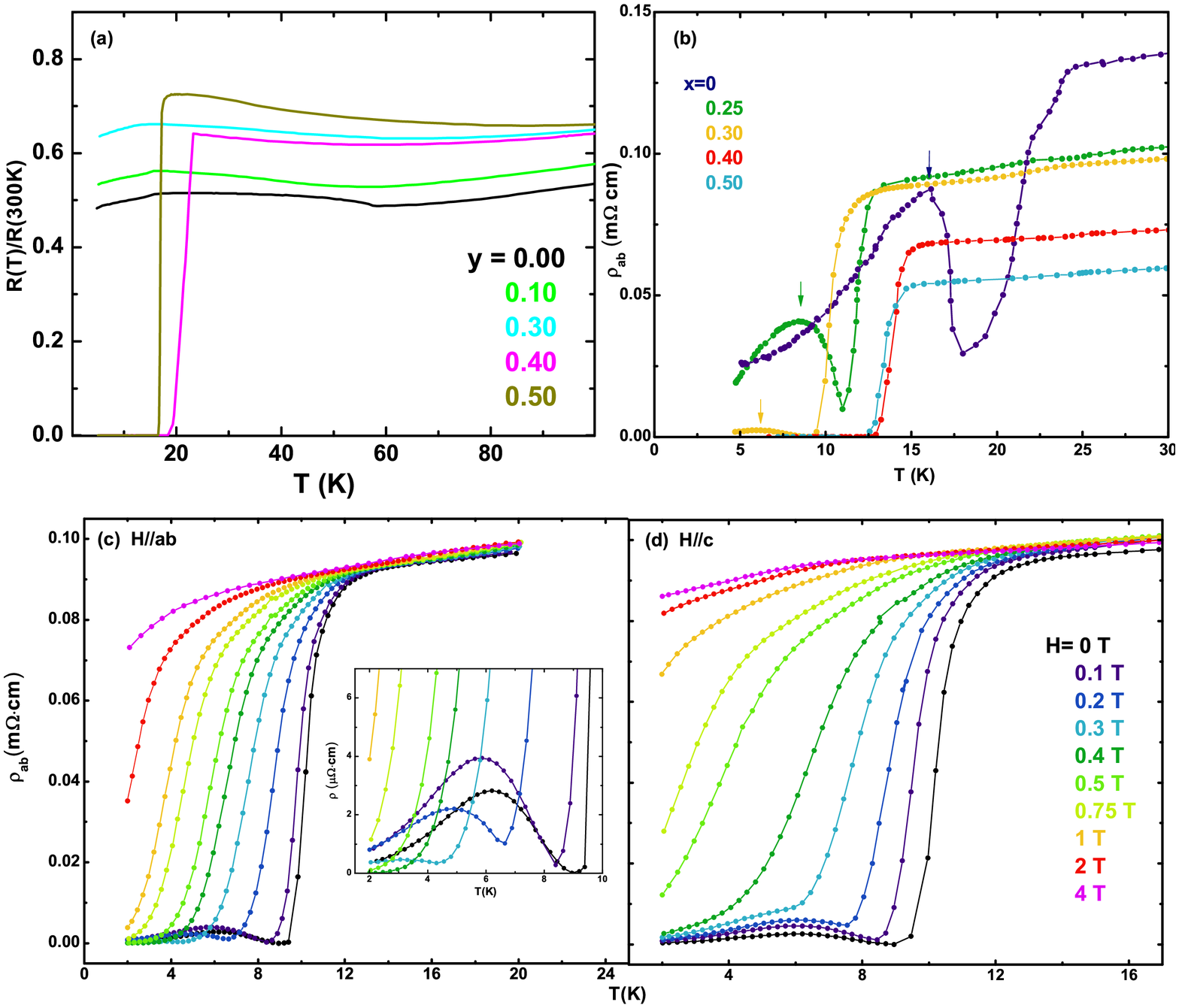}
\caption{(Color online). (a): Temperature dependent resistivity in
single crystals of Eu$_{1-y}$Ba$_{y}$Fe$_{1.8}$Co$_{0.2}$As$_2$.
(b): Temperature dependent resistivity in single crystals of
Eu$_{1-y}$Sr$_{y}$Fe$_{1.715}$Co$_{0.285}$As$_2$. (c) and (d):
Temperature dependent resistivity under different magnetic fields
for magnetically weakened
Eu$_{0.7}$Sr$_{0.3}$Fe$_{1.715}$Co$_{0.285}$As$_2$ single crystals,
where both superconducting transition and resistance reentrance
emerge at low temperatures. (c): H applied within ab-plane; (d): H
applied along c-axis.}\label{fig4}
\end{figure}

As reported by Wu et al.,\cite{Wu,Jiang} Eu$^{2+}$ ions in
EuFe$_2$As$_2$ would experience a metamagnetism from AFM to FM
induced by magnetic field at low temperature. In order to study how
superconductivity behaves together with AFM and FM order, we
measured the temperature dependent resistivity under different
magnetic field, specific heat and susceptibility of the optimal
doped sample EuFe$_{1.715}$Co$_{0.285}$As$_2$. As shown in Fig.3a,
the reentrance in resistivity is continuously suppressed both in
terms of intensity and corresponding temperature with increasing H
applied within ab-plane. When the H reaches to 1T, the resistivity
reentrance is completely suppressed, accompanying by a conventional
field-induced suppression on superconductivity. Figure 3b presents
the resistivity data measured with H applied along c-axis for the
same sample. Unlike the case in Fig.3a, the resistivity reentrance
does not exhibit conspicuous sign of shift or weakening with
increasing H, and the only change is the closing of the gap in
resistivity except for the expected down-shift of $T_c$ and
transition broadening induced by magnetic field. These results are
consistent with the results of susceptibility as shown in Fig.3c and
3d. Hence we may reckon that the easy axis lies in ab-plane, which
might have contributed to the insensitive behavior when applying H
along c-axis. Susceptibility and specific heat measurements further
confirm the antiferromagnetic transition at 17K. In Fig.3c, the Neel
temperature monotonously decreases with increasing H. The similar
suppression of the peak on specific heat could be observed in
Fig.3e. After H goes beyond 0.5 T, neither response from resistivity
nor susceptibility could be clearly observed. It is because a
metamagnetism from AFM to FM has taken place.\cite{Wu} These results
give strong evidence that a resistivity reentrance arises from the
antiferromagnetic ordering of $Eu^{2+}$ spins, indicating the
competition between antiferromagnetism (AFM) and superconductivity.
It is striking that the resistivity reentrance can be completely
suppressed by external magnetic field (H) due to the field-induced
metamagnetic transition from antiferromagnetism to ferromagnetism
for $Eu^{2+}$ spins. All above measurements concerning single
crystals of EuFe$_{2-x}$Co$_{x}$As$_2$ have contributed to the
picture that the spins of Eu$^{2+}$ ions tend to establish an
antiferromagnetic order around 17 K, and are easily tuned to
ferromagnetic order with small H. Such metamagnetism of Eu$^{2+}$
ions provides a good system to study the intriguing interaction
between AFM/FM and superconductivity. Antiferromagnetism appears to
have strongly counteracted superconductivity, while ferromagnetism
could coexist rather at ease with the superconductivity.

In order to make sure that it is AFM from Eu$^{2+}$ sublattice that
destroys superconductivity, we choose to partially substitute
Eu$^{2+}$ with nonmagnetic Ba$^{2+}$/Sr$^{2+}$. As expected,
superconductivity shows up at 23 K as shown in Fig.4a for the
Ba-doped crystals, being consistent with that in
BaFe$_{1.8}$Co$_{0.2}$As$_2$ crystal.\cite{Wangb} Figure 4b clearly
demonstrates the evolution in resistivity in
Eu$_{1-x}$Sr$_{x}$Fe$_{1.715}$Co$_{0.285}$As$_2$ system. With
increasing Sr-doping, the gap caused by reentrance is gradually
narrowed along with the suppression of the reentrance's peak. For
the crystal with x=0.3, superconducting transition is sharp, and the
resistivity reaches zero shortly before a resistivity reentrance
takes place. Further Sr-doping eventually kills the resistivity
reentrance, and stable superconducting phase can be achieved for the
crystal with x=0.5. Figure 4c and 4d present the temperature
dependent resistivity under different H applied within ab-plane and
along c-axis for the
Eu$_{0.7}$Sr$_{0.3}$Fe$_{1.715}$Co$_{0.285}$As$_2$ sample,
respectively. The resistivity reentrance can be continuously
suppressed with increasing magnetic field applied within ab-plane.
However, we didn't observe the same trend with H applied
perpendicular to ab-plane. This could probably be understood in
terms of the anisotropic magnetic structure and exchange integration
within or between Eu$^{2+}$ layers. In
$Eu_{1-x}Sr_xFe_{2-y}$Co$_{y}As_2$ system, the antiferromagnetism
appears to be more destructive to superconductivity, while the
ferromagnetism can coexist with superconductivity. It should be
pointed out that such metamagnetic transition from AFM to FM is
induced by a small external magnetic field. Therefore, the observed
behavior here is intrinsic.

In RNi$_2$B$_2$C system,  a pronounced resistivity reentrance was
observed when AFM transition in RC layers happens below
superconducting transition temperature\cite{Eisaki}, similar to the
observation in Eu$_{0.7}$Sr$_{0.3}$Fe$_{1.715}$Co$_{0.285}$As$_2$
system. If the AFM transition temperature T$_N$ is lower than
critical superconducting temperature T$_c$, T$_c$ could be
negatively scaled by R$^{3+}$ ions' de Gennes factor (often referred
to as DG factor) which quantifies the strength of local moment's
influence on conducting carriers\cite{Cho}. As suggested in s-wave
superconductors, the presence of local magnetic moment tends to
destabilize the bonding of spin singlet cooper pairs\cite{Eisaki}.
However, we find the linear DG scaling to be totally broken down in
our double-doped $Eu_{1-x}Sr_xFe_{2-y}$Co$_{y}As_2$ system. It
resembles the result in Dy-doped HoNi$_2$B$_2$C system\cite{Amici},
suggesting potential scattering from collective magnetic excitations
(magnons) rather than conventional long range magnetic order in
terms of RKKY model. According to the calculations, FM state is
always energetically favored than superconducting
state\cite{Greenside} at low temperature. On the other hand, if we
take into consideration possible emergence of vortices, the
advantageous energy state might change with respect to the density
of the vortices like in ErNi$_2$B$_2$C and TmNi$_2$B$_2$C
systems\cite{Canfield2, Gasser}. When T is much lower than T$_{FM}$,
the spontaneous vortex phase (SVP) appears more likely to win over
FM state, thus enabling superconductivity to survive with
ferromagnetism\cite{Greenside}. In EuFe$_{1.715}$Co$_{0.285}$As$_2$
system, T$_{FM}$ soon goes up over 22 K with increasing H, which
provides chances for SVP to be dominant and might offer one possible
explanation towards the unusual coexistence. Therefore, it should be
exciting to further adopt microscopic measurements like Small Angle
Neutron Scattering (SANS), Scanning Tunneling Microscopy (STM) and
magnetic Bitter-decoration to directly detect the surface condition
and magnetic structure of this system.

According to the theoretical work and experiments on
UGe$_2$\cite{Aoki}, the ground state for electron pairs in
ferromagnetic superconductor is usually spin-triplet rather than
conventional spin-singlet due to field-induced Fermi surface
splitting. Apart from the coexistence of superconductivity and
ferromagnetism in this system, we also observed moderately high
upper critical field for the
Eu$_{1-x}$Sr$_{x}$Fe$_{2-y}$Co$_{y}$As$_2$ system derived from Fig
4c, which exceeds the weak-coupling paramagnetic limit for
conventional superconductors\cite{Luk'yanchuk}. Given experimental
results above, we find it amazing that such iron-based
superconductor is likely to bear certain property of p-wave
superconductors, which seems to deviate a little from the widely
accepted \textbf{\emph{mixed s-wave}} configuration of iron arsenide
superconductors. Chances are that the magnetic Eu$^{2+}$ sublattice
exerts an additional impact on the symmetry of electronic wave
function, thus causing a stir among the conventional ground state
and excited states nearby.

In this Letter, we carefully measured the transport properties for
EuFe$_{2-x}$Co$_{x}$As$_2$ and
Eu$_{1-x}$Sr$_{x}$Fe$_{2-y}$Co$_{y}$As$_2$ single crystals.
Co-doping suppresses the SDW transition and induces superconducting
transition. In contrast to the case of BaFe$_{2-x}$Co$_{x}$As$_2$
system, a resistivity reentrance is observed at the temperature
corresponding to the antiferromagnetic ordering of $Eu^{2+}$ ions
and no zero resistivity is achieved in EuFe$_{2-x}$Co$_{x}$As$_2$
system. The resistivity reentrance can be completely suppressed by H
due to metamagnetic transition from antiferromagnetism to
ferromagnetism for $Eu^{2+}$ spins induced by H. These results
suggest that the antiferromagnetism destroys the superconductivity,
while the ferromagnetism can coexist with the superconductivity. The
coexistence of superconductivity and ferromagnetism denotes curious
property of p-wave superconductors.

We thank professor Lixin He and Yuwei Cui for helpful discussion in
terms of first principle calculation. This work is supported by the
Nature Science Foundation of China and by the Ministry of Science
and Technology of China (973 project No: 2006CB601001) and by
National Basic Research Program of China (2006CB922005).

\end{document}